\documentclass[fleqn,10pt]{wlscirep}
\usepackage[utf8]{inputenc}
\usepackage[T1]{fontenc}
\title{Synthesis, disorder and Ising anisotropy in a new spin liquid candidate PrMgAl$_{11}$O$_{19}$}

\author[1,2]{Yantao Cao}
\author[2]{Huanpeng Bu}
\author[2]{Zhendong Fu}
\author[2,3]{Jinkui Zhao}
\author[4,*]{Jason S. Gardner}
\author[5]{Zhongwen Ouyang}
\author[5,$\dag$]{Zhaoming Tian}
\author[1,$\ddag$]{Zhiwei Li}
\author[2,$\S$]{Hanjie Guo}

\affil[1]{Key Lab for Magnetism and Magnetic Materials of the Ministry of Education, Lanzhou University, Lanzhou 730000, China}
\affil[2]{Songshan Lake Materials Laboratory, Dongguan, Guangdong 523808, China}
\affil[3]{Institute of Physics, Chinese Academy of Sciences, Beijing 100190, China}
\affil[4]{Materials Science and Technology Division, Oak Ridge National Laboratory, Oak Ridge, Tennessee 37831, USA}
\affil[5]{School of Physics and Wuhan National High Magnetic Field Center,
Huazhong University of Science and Technology, Wuhan 430074, P. R. China.}
\affil[*]{gardnerjs@ornl.gov}
\affil[$\dag$]{tianzhaoming@hust.edu.cn}
\affil[$\ddag$]{zweili@lzu.edu.cn}
\affil[$\S$]{hjguo@sslab.org.cn}

\usepackage{multirow}
\usepackage{tabularx}
\usepackage{threeparttable}

\newcommand{\lmao}{LaMgAl$_{11}$O$_{19}$}
\newcommand{\pmao}{PrMgAl$_{11}$O$_{19}$}
\newcommand{\pzao}{PrZnAl$_{11}$O$_{19}$}

\begin{abstract}
  Here we report the successful synthesis of large single crystals of triangular frustrated \pmao\ using the optical floating zone technique. Single crystal X-ray diffraction measurements unveiled the presence of quenched disorder within the mirror plane, specifically $\sim$7\% of Pr ions deviating from the ideal 2\textit{d} site towards the 6\textit{h} site. Magnetic susceptibility measurements revealed an Ising anisotropy with the \textit{c}-axis being the easy axis. Despite a large spin-spin interaction that develops below $\sim$10~K and considerable site disorder, the spins do not order or freeze down to at least 50 mK. The availability of large single crystals offers a distinct opportunity to investigate the exotic magnetic state on a triangular lattice with an easy axis out of the plane.
  \\

  {\bf Keywords:} spin liquid, disorder, frustration, triangular lattice, Ising anisotropy
\end{abstract}
\begin{document}

\flushbottom
\maketitle
\thispagestyle{empty}

%\noindent Please note: Abbreviations should be introduced at the first mention in the main text ¨C no abbreviations lists. Suggested structure of main text (not enforced) is provided below.

\section*{Introduction}

Quantum spin liquids (QSLs) represent an intriguing state where the spins remain disordered even at zero Kelvin due to quantum fluctuations, albeit with strong spin-spin couplings \cite{Balents2010}. Achieving a QSL ground state is challenging because of the propensity for the spin sublattice to freeze as the temperature is lowered, especially around defects and/or disorder which act as pinning centers.  For example, in the triangular lattice of YbMgGaO$_4$, the presence of site disorder between the nonmagnetic ions Mg and Ga induces a spin glass behavior \cite{Ma2018}. However, disorder is not always detrimental to the QSL state. Studies on the pyrochlore oxides \cite{Savary-2017,Wen-2017} and 1\textit{T}-TaS$_2$ \cite{Murayama-2020} reveal that the quenched disorder does not compete, but rather cooperates with the frustration to induce strong quantum fluctuations, and may give rise to emergent spin disordered state responsible for the gapless excitations.

From an experimental point of view, a QSL state does not break any symmetry making it arduous to identify using a single technique \cite{Gao-2020}. Inelastic neutron scattering plays an important role in providing crucial evidence for a QSL state, as fractional excitations, such as spinons, manifest as a distinctive excitation continuum in the spectrum \cite{Knolle-2019}. The availability of high-quality large single crystals is essential for neutron scattering due to the small scattering cross-section and low neutron flux. Similar requisites hold true for other measurements, such as spin thermal transport, to minimize any grain boundary effects as well as the anisotropic characterizations of magnetic behaviors \cite{Yamashita2010,Ni2019}.

Here we present the single crystal growth of a hexaaluminate, \pmao, using the optical floating zone technique. Crystal growth and basic magnetic property measurements on this series of lanthanide aluminates have been reported\cite{Saber1985,Kahn-1981},
but a detailed characterization down to milikelvin is still lacking. Our previous results on the isostructural, polycrystalline PrZnAl$_{11}$O$_{19}$ already suggested the magnetic sublattice of Pr$^{3+}$ ions, decorating a triangular lattice, has the potential to host a Dirac QSL \cite{Ashtar2019,Bu-2022}. The triangular network of Pr ions reside within the \textit{ab} plane, and are connected via intermediate O ions which are also within the triangular plane, forming a nearly linear Pr-O-Pr bond with an angle of about 176.7$^{\circ}$. These layers are separated along the \textit{c}-axis by $\sim$11 \AA\ while the nearest neighbor Pr-Pr bond is  $\sim$5.59 \AA;  see Fig. \ref{structure}. As a comparison, the distance between triangular layers in YbMgGaO$_4$ are about 8 \AA\ with a nearest neighbor bond length of $\sim$3.40 \AA\ \cite{Li2019}. Moreover, the large difference in the ionic radii between the magnetic and nonmagnetic ions inhibits any site mixing between the magnetic and nonmagnetic ions. These structural features indicate that this system is an ideal quasi-two-dimensional system free from chemical disorder,  and great starting conditions for realizing the QSL state. Indeed, the signature of forming a QSL state has been observed in ac susceptibility measurements of PrZnAl$_{11}$O$_{19}$ revealing no spin freezing or ordering down to 50 mK, despite a considerable antiferromagnetic coupling strength of about -9 K. In addition, inelastic neutron scattering exhibits an abnormal broadening of low energy excitations at $\sim$1.5 meV \cite{Bu-2022}. Unfortunately single crystals of PrZnAl$_{11}$O$_{19}$ are currently not available at the size needed for inelastic neutron scattering.

By substituting Zn with the less evaporative Mg element, we have succeeded in growing sizable single crystals of \pmao\ suitable for neutron scattering measurement. However, we have identified site disorder within the mirror plane, with about 7\% of the Pr ions displaced from the ideal position. Magnetic susceptibility measurements show the moments lying perfectly along the \textit{c}-axis (or perpendicular to the triangular plane)  and do not freeze down to 50 mK. We argue that the system keeps fluctuating despite strong couplings and site disorder, which makes this system unique for a triangular system with Ising character.

\section{Methods}

Polycrystalline samples were prepared using a standard solid-state reaction technique. Raw materials of Pr$_6$O$_{11}$ (99.99\%), MgO(99.99\%), and Al$_2$O$_3$ (99.99\%) were dried at 900$^\circ$C over night prior to reaction to avoid moisture contamination. Then, stoichiometric amounts of the raw materials were mixed and ground thoroughly, pressed into pellets and sintered at 1400$^\circ$C $\sim$ 1600$^\circ$C with several intermediate grindings. The powder sample was mixed with about  1\% $\sim$ 2\% excess of MgO and pressed into a cylindrical rod of $\sim$6 mm in diameter and $\sim$140 mm ($\sim$35 mm) in length as a feed (seed) rod using a hydrostatic pressure of 70 MPa. The obtained feed and seed rods were sintered at 1500$^\circ$C for 2 hours. Subsequent single crystal growth was conducted in an optical floating zone furnace (HKZ300) in pure argon atmosphere at 10 bar.  After floating zone growth, the several-centimeter-sized as-grown crystal was annealed at 1000$^\circ$C in flowing O$_2$ for 24 hours and then slowly cooled down to room temperature in order to avoid any possible oxygen vacancy.

The single crystal X-ray diffraction (XRD) measurement was performed on a XtaLAB Synergy diffractometer (Rigaku) at room temperature using the Mo-$K_\alpha$ radiation. The experimental conditions are tabulated in Tab. \ref{SXRD1}. Part of the single crystal was crushed into powders for powder XRD measurement performed on a MiniFlex diffractometer (Rigaku) with the Cu-$K_\alpha$ radiation. JANA \cite{Jana} and FULLPROF \cite{Fullprof} packages were used for crystal structure refinement.

Heat capacity measurements were carried out on the Physical Property Measurement System (PPMS, Quantum Design) equipped with a dilution insert using the relaxation method. The DC magnetic susceptibility between 2 and 350 K was measured using the vibrating sample magnetometer (VSM) option of the PPMS. The AC magnetic susceptibility
between 0.05 and 15 K was measured using the ACMS-II and ACDR options of the PPMS. For the AC
susceptibility measurement, a driven field of 1-3 Oe in amplitude was used.

\section{Results}

\begin{figure}
  \centering
  \includegraphics[width=0.8\columnwidth]{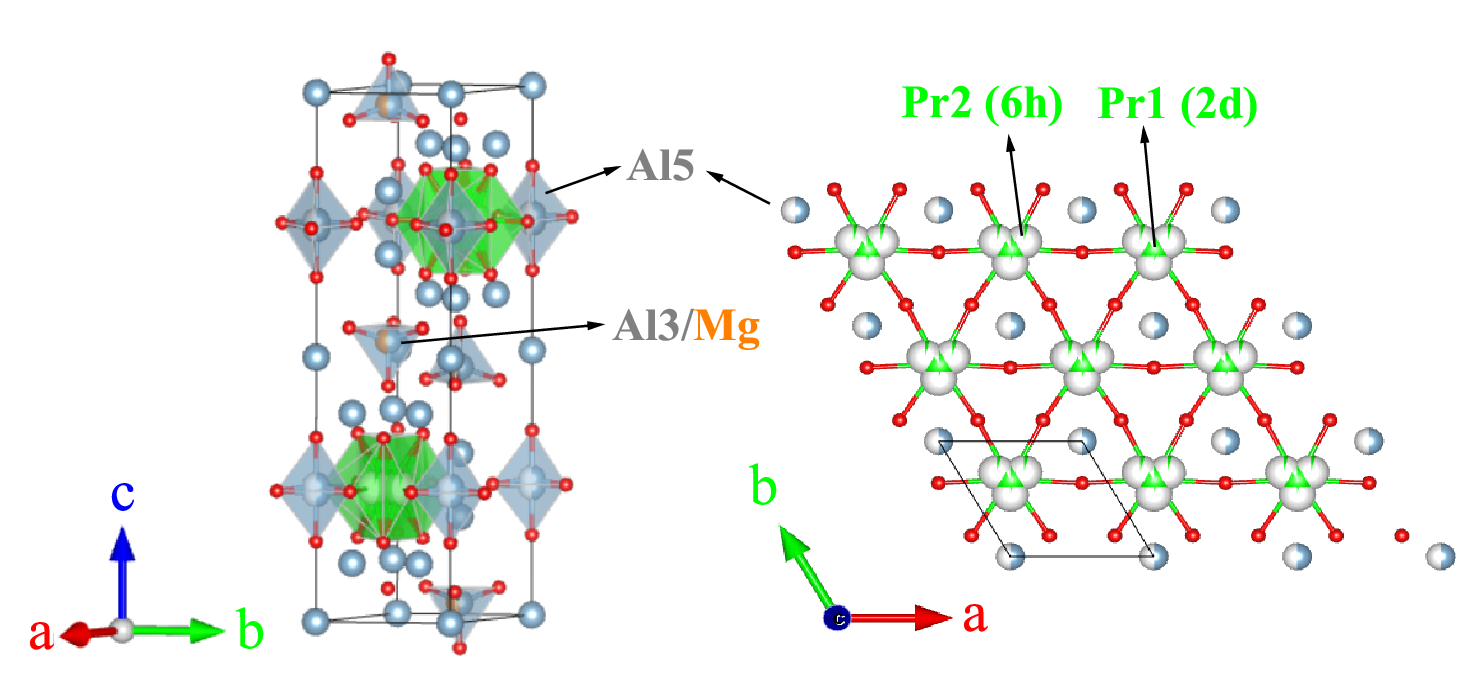}\\
  \caption{The crystal structure of \pmao\ extracted from single-crystal XRD refinement. Green - Pr; grey - Al; red - O; orange - Mg. The mirror plane including the disordered Pr ions is depicted in the right panel.}\label{structure}
\end{figure}

\begin{figure}
  \centering
  \includegraphics[width=0.8\columnwidth]{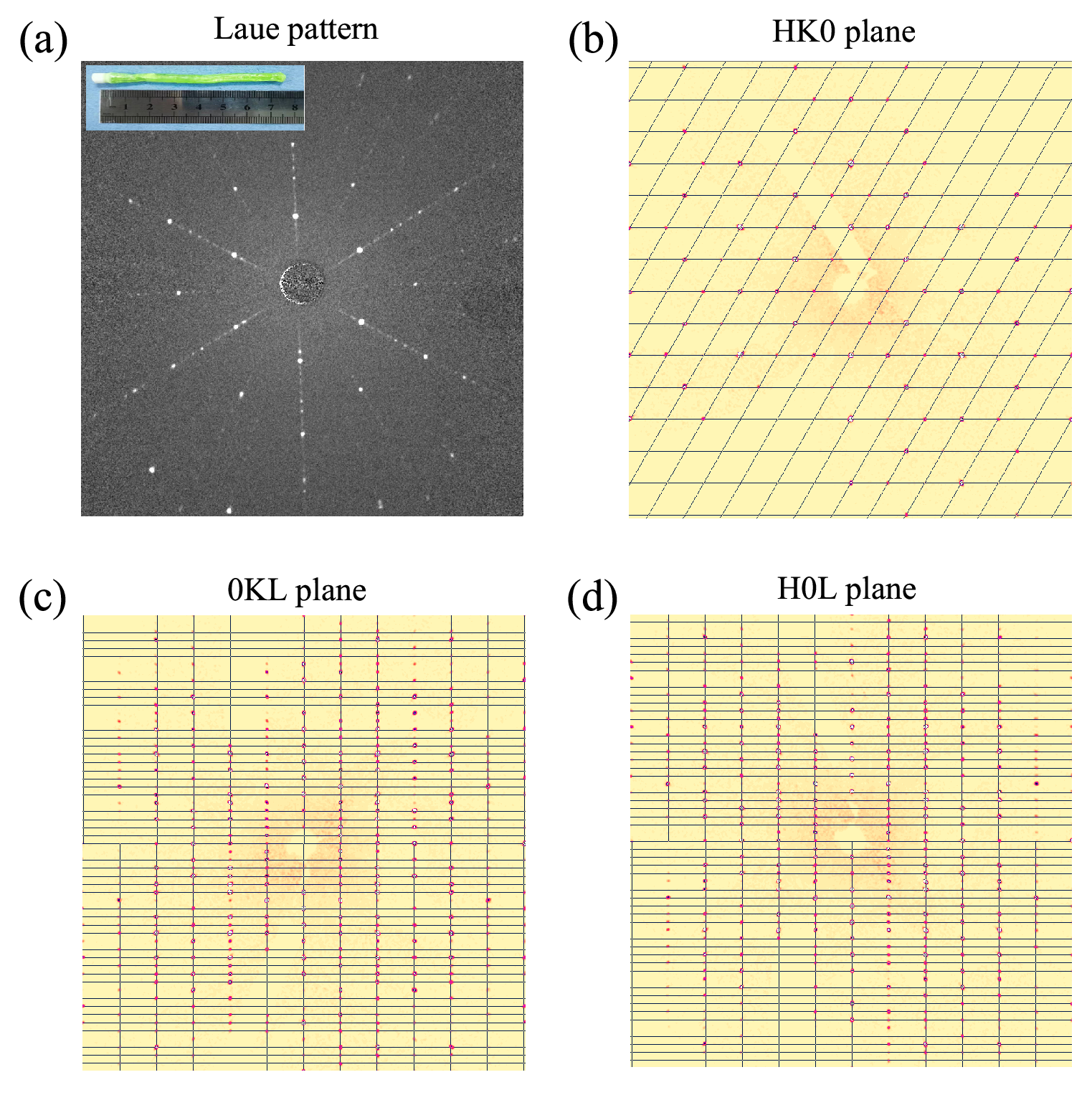}\\
  \caption{X-ray studies of PrMgAl$_{11}$O$_{19}$.  (a) A Laue pattern with the X-ray beam approximately along the \textit{c} axis. The inset shows a photo of the as-grown single crystal. (b-d) Precession images within the HK0, 0KL and H0L planes constructed from the single-crystal XRD measurements.}\label{XRD}
\end{figure}

The single crystal growth was initially attempted in a pure oxygen atmosphere, which turns out to be unstable and results in cavities along the as-grown crystals. This occurrence indicates the dissolution of oxygen in the melt, forming stable bubbles during the growth process. Therefore, subsequent growths were performed in a pure Ar atmosphere at 10 bar pressure in order to minimize the evaporation of the Mg element. The upper and lower rods were counter-rotated at a rate of 20 rpm, and moved downwards simultaneously at a rate of 2 mm/h. These conditions led to the successful growth of centimeter-sized single crystals, as depicted in the inset of Fig. \ref{XRD}(a). The single-domain nature was confirmed by Laue diffraction measurements along the rod. A typical Laue pattern is shown in Fig. \ref{XRD}(a) with the X-ray beam nearly parallel to the \textit{c}-axis. Laue measurements also confirm that the crystal was grown along the \textit{a}* direction.

The quality of the single crystal was further characterized by X-ray diffraction of single crystals and crushed crystals. All peaks in the powder diffraction pattern can be indexed by the space group $P6_3/mmc$, indicating an impurity-free phase; see the Supplementary Materials (SM) \cite{supp}. The precession images within the HK0, 0KL, and H0L planes from the single crystal XRD measurements are shown in Fig. \ref{XRD}(b-d). The sharp peaks confirm the high quality of the crystal. The structure refinement is consistent with the findings reported by Kahn \textit{et al}. for LaMgAl$_{11}$O$_{19}$ \cite{Kahn-1981}.
The final structure parameters extracted from the single crystal refinement are listed in Tab. \ref{SXRD2}. It was found that about 7\% of the Pr ions at the 2\textit{d} (Pr1) site were shifted about 0.6 \AA~ away from the ideal position towards the 6\textit{h} (Pr2) site. Moreover, within the mirror plane of the structure, the Al5 ion displayed an anomalous atomic displacement parameter (ADP) along the \textit{c}-axis, $U_{33}$, which amounts to 0.018 \AA$^2$, compared to the others of around 0.007 \AA$^2$. Therefore, it is set to be slightly off the mirror plane with half occupancy at the 4\textit{e} site. Note, that with the resolution of our lab-based powder X-ray diffractometer, this anomalous ADP would be within errors and not highlighted.
The last unsolved issue is the occupied position of the Mg ions. It is difficult to distinguish Mg$^{2+}$ from Al$^{3+}$ by means of X-ray diffraction because of the same number of electrons for both ions. However, neutron powder diffraction measurement on its sister compound CeMgAl$_{11}$O$_{19}$ \cite{Ce} indicated that Mg is likely positioned at the Al3 site with an AlO$_4$ tetrahedron coordination, which also shows an anomalous ADP if solely occupied by Al ions. Therefore, the Mg ions are mixed with the Al3 ion at the 4\textit{f} site with 0.5 occupancy. The refined structure is shown in Fig. \ref{structure} and those particular ions are highlighted.

\begin{table}
\caption{Experimental conditions for the single crystal XRD measurements, and agreement factors for the refinement.\label{SXRD1}}
%\begin{ruledtabular}
\begin{tabularx}{\linewidth}{l r}%{lllll}
%\toprule
\hline
\hline
Formula                                                      &    PrMgAl$_{11}$O$_{19}$ \\
Space group                                                  &    $P6_3/mm\/c$ (No. 194)\\
\textit{a, b} ($\textrm{\AA}$)                               &    5.58700(10) \\
\textit{c} ($\textrm{\AA}$)                                  &    21.8732(6)  \\
\textit{V} ($\textrm{\AA}^3$)                                &    591.29(2) \\
\textit{Z}                                                   &    2 \\
2$\Theta$ ($^\circ$)                                         &    3.72 - 82.16 \\
No. of reflections, $R_\mathrm{int}$                         &    12534, 4.11\%\\
No. of independent reflections                               &    614 \\
No. of parameters                                            &    48 \\
\multirow{3}*{Index ranges}                                  &    -9 $\leq$ \textit{H} $\leq$ 10, \\
                                                             &    -10 $\leq$ \textit{K} $\leq$ 10,  \\
                                                             &    -39 $\leq$ \textit{L} $\leq$ 39 \\
\textit{R}, \textit{wR$_2$}   $^*$\tnote{1}                     &    1.69\%, 4.60\% \\
Goodness of fit on $F^2$                                     &    1.42\\
Largest difference peak/hole (\textit{e}/$\textrm{\AA}^{3}$) &    0.34/-0.38\\
\hline
\hline
\end{tabularx}
\begin{tablenotes}
\item[1] $^*$For a direct visualization of the fitting quality, see $F_{obs.}^2$ vs. $F_{cal.}^2$ in the SM \cite{supp}.
\end{tablenotes}
\end{table}

\begin{table}
\caption{Refined crystal structure parameters.\label{SXRD2}}
\begin{tabularx}{\linewidth}{X X X X X}
%\begin{tabular}{lllll}
\hline
\hline
Atom  & occ. &  x & y & z  \\
Pr1 (2d)  &  0.928(10)   & 1/3 & 2/3  & 0.75  \\
Pr2 (6h)  &  0.024(3)   & 0.271(5) & 0.729(5) & 0.75  \\
Al1 (2a)  & 1   &  0   &  0  & 0           \\
Al2 (4f)  &  1   & 1/3 & 2/3 & 0.18998(4)   \\
Al3 (4f)  & 0.5 &  1/3 & 2/3 & 0.47275(4) \\
Mg  (4f)  & 0.5 &  1/3 & 2/3 & 0.47275(4) \\
Al4 (12k) &  1   & 0.16738(4) & 0.33477(9) &0.60839(2) \\
Al5 (4e)  & 0.5 &  0   &  0  & 0.2428(3) \\
O1  (6h)  & 1   &  0.18122(15) & 0.3624(3)& 0.25  \\
O2  (12k) & 1   & 0.15214(11) & 0.3043(2) &  0.44609(6) \\
O3  (12k) & 1   & -0.0112(2) & 0.49442(11) & 0.34808(5) \\
O4  (4f)  & 1   &  1/3  & 2/3  &  0.55837(10) \\
O5  (4e)  & 1   &  0      & 0 & 0.34818(9) \\
\end{tabularx}

\begin{tabularx}{\linewidth}{X X X X }
\hline
Atom  & $U_{11}$/$U_{12}$ (\AA$^2$) & $U_{22}$/$U_{13}$ (\AA$^2$) & $U_{33}$/$U_{23}$ (\AA$^2$)\\
Pr1   &0.0099(3) & 0.0099(3) & 0.00756(13) \\
      & 0.00493(15) & 0& 0 \\
Pr2   & 0.030(5) & 0.030(5) & 0.019(4)\\
      & 0.007(5) & 0& 0 \\
Al1   & 0.0070(3) & 0.0070(3) & 0.0078(5) \\
      & 0.00352(14) & 0  & 0 \\
Al2   & 0.0072(2) & 0.0072(2)  & 0.0067(4) \\
      & 0.00358(11) & 0& 0 \\
Al3/Mg   & 0.0068(2) & 0.0068(2) & 0.0076(4) \\
      & 0.00340(12) & 0& 0 \\
Al4   & 0.00682(17) & 0.0071(2) & 0.0075(2) \\
      & 0.00353(11) & -0.00007(7) & -0.00013(14)\\
Al5   & 0.0069(3) & 0.0069(3) & 0.018(3) \\
      & 0.00347(17) & 0 & 0 \\
O1    & 0.0107(5) & 0.0075(6) &  0.0080(7)\\
      & 0.0038(3) & 0 & 0 \\
O2    & 0.0096(4) & 0.0123(5) & 0.0091(5) \\
       & 0.0062(2)  & 0.00117(19) & 0.0023(4) \\
O3    & 0.0079(4) & 0.0077(3) & 0.0086(5) \\
      & 0.0040(2) & 0.0005(3) & 0.00024(17) \\
O4    & 0.0075(5) & 0.0075(5) & 0.0119(9)\\
      & 0.0037(2) & 0& 0 \\
O5    & 0.0078(4) & 0.0078(4) & 0.0103(8)\\
     &0.0039(2) & 0& 0 \\
\hline
\hline
%\toprule
\end{tabularx}
\end{table}

\begin{figure}
  \centering
  \includegraphics[width=0.8\columnwidth]{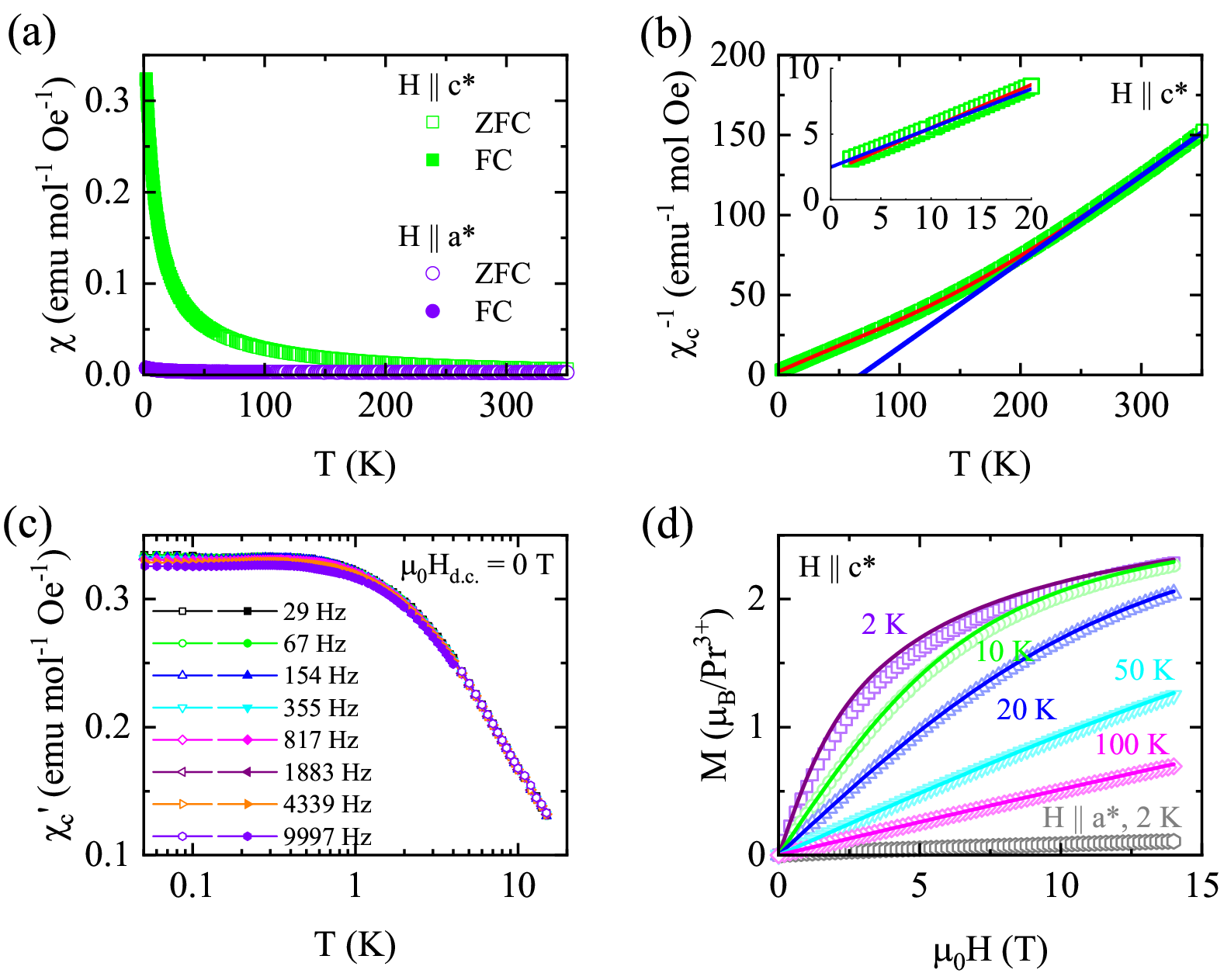}\\
  \caption{Temperature and field dependence of the dynamic susceptibility of PrMgAl$_{11}$O$_{19}$. (a) Temperature dependence of the magnetic susceptibility measured with the magnetic fields applied along the reciprocal \textit{a}* and \textit{c}* directions. (b) Temperature dependence of the inverse susceptibility, $\chi^{-1}_c(T)$. The red and blue curves are according to the CEF fit and Curie-Weiss fit, respectively. The inset highlights the low-temperature region. (c) Temperature dependence of the real component of the ac susceptibility, $\chi'_c$(T), measured at various frequencies. (d) Isothermal magnetization measured at various temperatures. The solid lines represent the magnetization calculated from the CEF parameters.}\label{sus}
\end{figure}

Moving to the magnetic property measurements. The X-ray photoelectron spectroscopy (XPS) measurements indicate a pure Pr$^{3+}$ state \cite{supp}. Fig. \ref{sus}(a) shows the temperature dependence of the magnetic susceptibility measured along different crystallographic directions. Note, that the \textit{c-} and \textit{c}*-directions are identical. The data shows pronounced anisotropy behavior with the moment predominantly aligned along the \textit{c}-axis. No difference between the zero-field-cooled (ZFC) and field-cooled (FC) curves can be observed, consistent with a dynamical (paramagnetic) state. The temperature dependence of $\chi^{-1}_c$ is shown in Fig. \ref{sus}(b).  A Curie-Weiss (CW) fit, $\chi^{-1}_c = (T-\theta_{CW})/C$, to the data above 250 K yields a positive Curie temperature $\theta_{CW}$ of 67 K. It should be noted that at these elevated temperatures we expect a large contribution from the crystal electric field (CEF) effect since two CEF levels were observed at $\sim$12 and 35 meV \cite{Bu-2022} in the isostructural PrZnAl$_{11}$O$_{19}$.  At low temperatures where only the low-lying CEF ground state is occupied, the influence from the higher excitation levels is negligible, a CW-like fit then provides a measure of the intersite interactions. A fit below 10 K yields $\theta_{CW}$ of -8 K, indicating strong antiferromagnetic couplings; see Fig. \ref{sus}(b). The effective moment is 5.18 $\mu_B$/Pr, which is close to the expected value of $g_c^{ESR}\sqrt{(S(S+1)} = 4.42 \mu_B$/Pr, where $g_c^{ESR}$ = 5.1 is the \textit{g} factor determined from ESR measurement \cite{supp}, and \textit{S} = 1/2, implying that the low temperature magnetic properties are governed by an effective \textit{S} = 1/2 state.
AC susceptibility measurements along the $c$-axis show no sign of long-range ordering, nor any spin freezing down to 50 mK; see Fig. \ref{sus}(c).
The field dependence of isothermal magnetization (Fig. \ref{sus}(d)) reveals negligible magnetization along the \textit{a}* direction compared to that of the \textit{c} direction. It is about 20 times larger along the \textit{c} axis at 2 K and 14 T, which can be easily explained by a slight misalignment  (~2 degrees) of the crystal during the measurement. Thus, the system can be considered as a well-defined Ising system.

In order to have more insights into the magnetic ground state, we performed CEF analysis and fit the CEF Hamiltonian to $\chi_c^{-1}(T)$. Although there are two inequivalent Pr sites, most of the Pr ions ($>$ 90\%) are still located at the $2d$ site with a $D_{3h}$ symmetry. Thus, we start with a model disregarding the disorder effect at the Pr site. In such a case, the CEF Hamiltonian can be expressed as
\begin{equation}\label{CEF}
  \mathcal{H}_{CEF} = \sum_{n,m}B_n^mO_n^m,
\end{equation}
where $O_n^m$ are the Stevens operators, and only the CEF parameters $B_2^0$, $B_4^0$, $B_6^0$ and $B_6^6$ are nonzero. The presence of the $O_6^6$ term admixes those states $|J,m_J\rangle$ differ in $m_J$ by 6. For Pr$^{3+}$, the total angular momentum number \textit{J} = 4. Thus, there are three doublets: two are formed by a linear combination of $|\pm4\rangle$ and $|\mp2\rangle$, and the other is $|\pm1\rangle$.
The three singlets are consisting of $|3^a\rangle = 1/\sqrt2(|3\rangle + |-3\rangle)$, $|3^b\rangle = 1/\sqrt2(|3\rangle - |-3\rangle)$, and $|0\rangle$. Therefore, if the ground state is a doublet and well separated from the first excited state, it will be highly anisotropic since $\langle J^{\pm}\rangle$ = 0. To estimate the energy separation, we use the PyCrystalField package \cite{Scheie} to fit $\mathcal{H}_{CEF}$ to the magnetic susceptibility data. To do so, we calculate the magnetic susceptibility $\chi_\mathrm{CEF}^c = M_{CEF}^c/H$ based on the CEF Hamiltonian, and treat the influence of the surrounding ions to a mean field level, so that $\chi_\mathrm{cal}^c = \chi_\mathrm{CEF}^c/(1-\lambda\chi_\mathrm{CEF}^c)$.
By minimizing $\chi^2 = \sum_i (1/\chi_{\mathrm{cal},i}^c - 1/\chi_{\mathrm{obs},i}^c)^2$, the best fit yields $B_2^0$ = -0.84181 meV, $B_4^0$ = -0.00516 meV, $B_6^0$ = -0.00012 meV, $B_6^6$ = -0.00602 meV, and $\lambda$ = -1.19 T/$\mu_B$. The eigenvectors and eigenenergies can be found in the SM \cite{supp}.
The negative $\lambda$ indicates an antiferromagnetic interaction among the spins. From $\lambda = zJ_c/(\mu_Bg_J)^2$, where \textit{z} is the number of nearest neighbors, the exchange constant $J_c$ can be estimated as $\sim$ -0.73$\times$10$^{-2}$ meV (-0.08 K). The small exchange constant is partly due to the large effective moment of Pr$^{3+}$ (\textit{J} = 4) \cite{Scheie2020}. In the effective spin \textit{S} = 1/2 picture, it will be enhanced by a factor of $\frac{J(J+1)}{S(S+1)}$ = 26.7 ($\sim$ -2.1 K). Correspondingly, the Curie-Weiss temperature at low temperature can be estimated as $\theta_{CW}^{\mathrm{cal.}} = S(S+1)zJ_c/3k_B \sim$ -3 K, which is roughly consistent with the experimental value. The calculated $(\chi_\mathrm{cal}^c)^{-1}$ is shown in Fig. \ref{sus}(b), which reproduces the overall data very well. Using the CEF parameters and $\lambda$, the calculated isothermal magnetization is shown in Fig. \ref{sus}(d). Again, the calculated curves agree well with the experimental values. The small discrepancy below $\sim$10 K may suggest the onset of spin-spin interactions. Moreover, the \textit{g} factors corresponding to the ground state doublets amount to $g_{ab} = g_J\langle\mp|J^\mp|\pm\rangle$ = 0 and $g_c = 2g_J\langle\pm|J_z|\pm\rangle$ = 5.6, consistent with the Ising character as inferred from the susceptibility measurements. The calculated $g_c$ is also close to the ESR result \cite{supp}. These good agreements indicate that the magnetic ground state is well captured by our CEF analysis, although the excited state may be overestimated.

%The isothermal magnetization along \textit{c}-axis are well described by the Brillouin function at high temperatures \cite{Ashcroft}. As the temperature is lowered, the M versus H curve starts to deviate from the Brillouin function around 20~K, demonstrating the onset of intersite spin-spin correlations although these do not become long-range even at 50 mK.

\begin{figure}
  \centering
  \includegraphics[width=0.8\columnwidth]{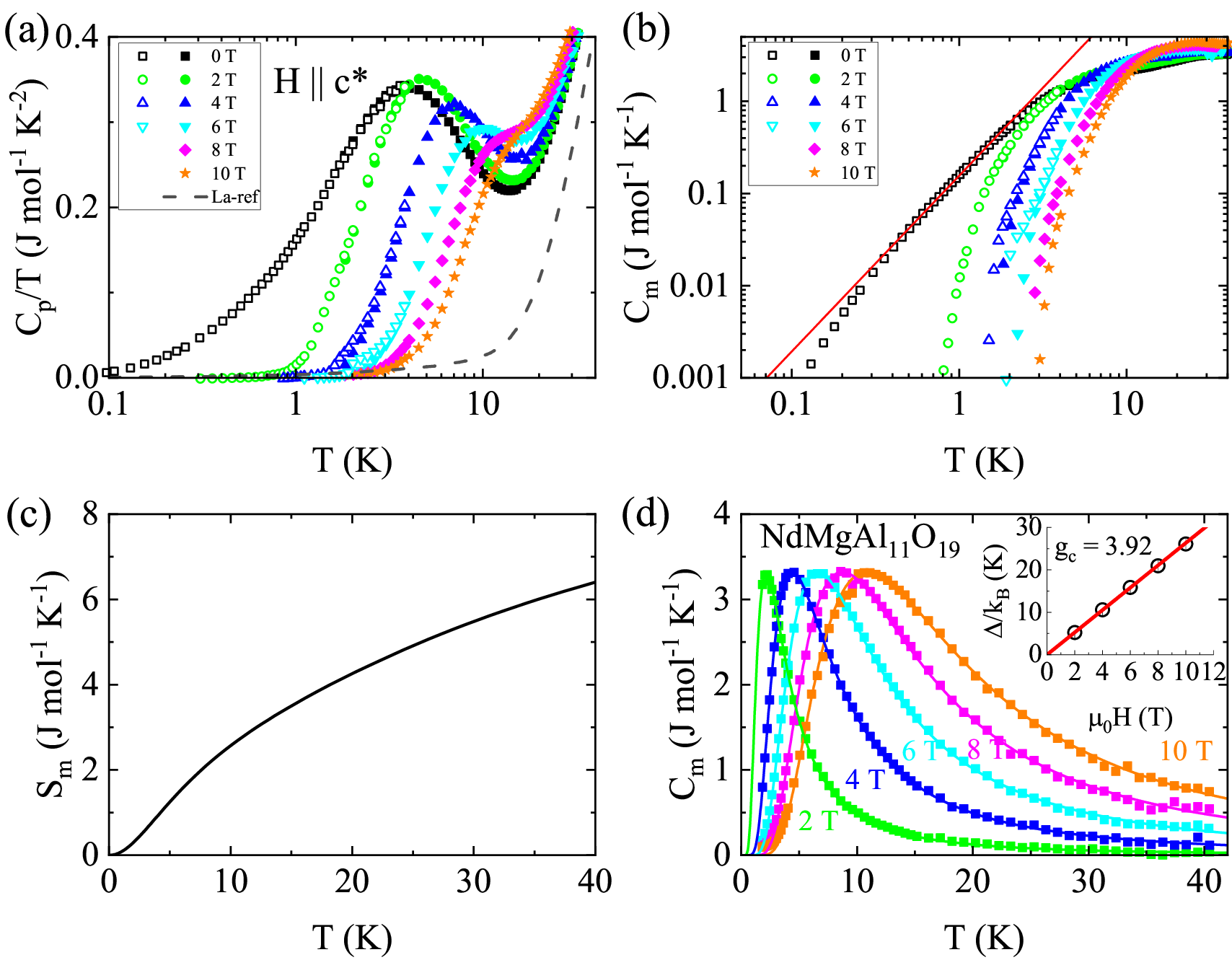}\\
  \caption{Heat capacity from hexaaluminate single crystals. (a) $C_p/T$ versus \textit{T} measured under various magnetic fields. The dashed line represents the phonon contribution obtained from the nonmagnetic counterpart \lmao. (b) The magnetic specific heat is obtained by subtracting the phonon contributions (\lmao) from the data of PrMgAl$_{11}$O$_{19}$. The red line represents a fit, $C_m = AT^{\alpha}$, to the zero field data in the temperature range between 0.2 and 2 K, with $\alpha$ = 1.91(1). (c) Representative change of the magnetic entropy in zero field. (d) Temperature dependence of the magnetic specific heat for NdMgAl$_{11}$O$_{19}$ together with a two-level Schottky fit. The inset shows the evolution of the gap as a function of the applied magnetic fields. A fit of $\Delta = g_c \mu_B H$ to the gap yields $g_c$ of 3.92.}\label{HC}
\end{figure}

Figure \ref{HC}(a) shows the specific heat measurements for \pmao\ under various magnetic fields. A nonmagnetic sample LaMgAl$_{11}$O$_{19}$ was also measured as a comparison. A broad hump can be observed at $\sim$4 K in the zero field data. With an increase in magnetic field, the peak is shifted towards high temperatures, reminiscent of the Schottky anomaly. However, upon subtracting the phonon contributions using LaMgAl$_{11}$O$_{19}$ as a reference sample, the magnetic component, $C_m$, can be obtained and shown in Fig. \ref{HC}(b). It is apparent that $C_m$ of \pmao\ does not exhibit a typical (multilevel, broadened) Schottky behavior \cite{Schotte1975,Bredl1978}, especially at the high temperature side where it displays a continuous increase up to 40 K. As a comparison, Fig. \ref{HC}(d) shows the $C_m$ of NdMgAl$_{11}$O$_{19}$ which displays the prototypical Schottky signature of a 2-level system. Along with the experimental data, a two-level Schottky fit, $C_m = p\cdot R(\frac{\Delta}{k_BT})^2\frac{\mathrm{exp}(\Delta/k_BT)}{[1+\mathrm{exp}(\Delta/k_BT)]^2}$, where $R$ is the ideal gas constant, $\Delta$ is the gap between the two levels, $k_B$ is the Boltzmann's constant, and $p$ represents the fraction of free ions \cite{Gopal} is depicted. In polycrystalline \pzao, the $C_m$ follows a power law in the temperature range between 0.2 and 2 K \cite{Bu-2022}. However in our PrMgAl$_{11}$O$_{19}$ data (Fig. \ref{HC}(b)), $C_m$ decreases faster than anticipated by a power law, suggesting the presence of a gap not seen in the Zn-analogue. On the other hand, the released entropy increases monotonically with increasing temperature, as shown in Fig. \ref{HC}(c), in a similar manner to that seen in the polycrystalline PrZnAl$_{11}$O$_{19}$ sample \cite{Bu-2022}.

\section{Discussions}
One essential discovery in the present study is the quenched disorder observed from single crystal X-ray diffraction refinements. The role of disorder in QSL is still elusive, and varies significantly from sample to sample. While it may drive a system into a glassy state, it may also enhance quantum fluctuations and potentially facilitate the formation of a QSL state, as demonstrated both theoretically and experimentally \cite{Savary-2017,Wen-2017,Murayama-2020,Fukukawa-2015}. The AC susceptibility data in Fig. \ref{sus}(c) and the lack of frequency-dependent in the data down to  50 mK, indicate that the system is still dynamic even though there is measurable site disorder. One possible scenario is the strong fluctuations resulting from the geometric frustration of antiferromagnetically coupled, easy-axis spins on the triangular lattice cannot be relieved by the site disorder ($\sim$7\%), which would intuitively result in the destruction of the spin liquid state.
Alternatively, the site disorder may cooperate with the frustration and stabilize a spin-liquid-like state as proposed for YbMgGaO$_4$ \cite{Ma2018,Kimchi2018,Li2019-PRL,Wu2019,Ma2021,Li2020-JPMC}. The present study does not have sufficient information to distinguish between these possibilities. Systematic studies on samples with different degrees of disorder will be instructive to clarify this point.  Similar questions have been addressed in the pyrochlore oxide Yb$_2$Ti$_2$O$_7$, where a sharp peak in the specific heat for polycrystalline samples, due to a magnetic transition at $\sim$265 mK \cite{Hodges2002}, was not observed in many single crystals, where only a broad peak at lower temperature was observed \cite{Yaouanc2011,Ross2012}. To investigate the small degree of site disorder found here within powder samples will need higher spatial resolution measurements, such as neutron/X-ray pair distribution function (PDF) method and extended X-ray absorption fine structure (EXAFS) to illustrate the local structure differences.

Another prominent feature of \pmao\ is the distinct Ising anisotropy revealed by the magnetization measurements. While the Heisenberg model usually predicts a magnetically ordered state for a triangular antiferromagnet, the Ising model can result in a macroscopically degenerated spin liquid state \cite{Wannier-1950}. Recent examples illustrating this concept is the TmMgGaO$_4$ \cite{Shen2019,Li2020,Li2020-NC} and neodymium heptatantalate, NdTa$_{7}$O$_{19}$ \cite{Arh-2022}. However, TmMgGaO$_4$ shows a partial order state below 0.7 K, and a lack of single crystals for NdTa$_{7}$O$_{19}$ hinders further explorations such as inelastic neutron scattering, as well as the exotic magnetic behaviors related to the crystalline directions. In this sense, the availability of large single crystals for \pmao\ provides a promising opportunity to investigate the triangular Ising model in depth.

\section{Conclusions}

In summary, we have successfully synthesized centimeter-sized single crystals of a spin liquid candidate \pmao\ by means of the optical floating zone technique. Single crystal structure refinement unveiled the presence of about 7\% quenched disorder at the Pr site in our sample. Directional magnetization measurements show a well-defined out-of-plane Ising anisotropy, which can induce strong fluctuations at a triangular lattice, as confirmed by the ac susceptibility measurements down to 50 mK. The availability of single crystals for this compound paves the way to explore the exotic magnetic properties by means of neutron scattering in the future.

While under review we became aware of a parallel work by Z. Ma \textit{et al.}, who also conclude this sample is an Ising spin on the triangular lattice \cite{Ma2024}. Here, by combining the CEF calculation, magnetization data and estimated \textit{g}-factors by ESR results, we present more robust evidence on the Ising anisotropy. Moreover, without single crystal structure refinement, the disorder at the Pr site was not reported in that work.

\section{Future Perspectives}

While disorder or defect is inevitable in real materials, many researchers endeavour to produce crystals as perfect as possible to obtain their intrinsic properties, allowing one to validate theoretical models. On the other hand, in recent years, it has been realized that disorder may result in exotic phases such spin-liquid-like random-singlet state. Unveiling the role of disorder systematically turns out to be as challenging as producing ideal crystals. In the title compound, the site disorder is within the triangular magnetic sublattice, which is unique compared to, e.g. YbMgGaO$_4$ where the site disorder occurs completely at the nonmagnetic site. By substituting Pr with another rare earth element, we expect to find evidence of different degrees of site disorder at the magnetic site (presumably this is also true when the nonmagnetic ions are substituted too), which will be helpful in future studies to manipulate the disorder in a controllable way.  Substituting the rare-earth ion will also change the local spin character decorating the triangular lattice due to a change in the crystalline electric field scheme.  This provides researchers with another tuning parameter for the magnetism in these hexaaluminates and should result in the discovery of systems with the spins confined to the triangular plane or with more exotic spin textures.  As with the magnetic pyrochlore oxides \cite{Gardner2010}, the availability of large single crystals and the large number of chemical substitutions that present themselves, we envisage this avenue of research to be very fruitful, resulting in materials with a variety of properties including superconductivity, emergent quantum phenomena, exotic spin texture and other quantum spin liquids.

\section{Acknowledgments}

This research was funded by the Guangdong Basic and Applied Basic Research Foundation (Grant No. 2022B1515120020), and the NSF of China with Grant No. 12004270 and 11874158. A portion of this work was performed on the Steady High Magnetic Field Facilities, High Magnetic Field Laboratory, CAS. A portion of this work was supported by the Laboratory Directed Research and Development (LDRD) program of Oak Ridge National Laboratory, managed by UT-Battelle, LLC for the U.S. Department of Energy.

\section{Author contributions}

Conceptualization: H.G., Z.T.; crystal growth: Y.C. and H.G.; measurement: Y.C., H.B., Z.T. and Z.O.; analysis: Y.C. and H.G.; validation: Z.F., J.S.G., Z.L., and J.Z.; writing: J.S.G. Z.T. and H.G. with inputs from all authors. All authors have read and agreed to the published version of the manuscript.

\section{Conflict of interest}
The authors declare no conflict of interests.

\bibliography{ref}

\section{Supplementary Materials}

\subsection{Powder X-ray diffraction}

\renewcommand{\thefigure}{S\arabic{figure}}
\setcounter{figure}{0} 

\begin{figure}[htbp]
\centering
\includegraphics[width=0.5\textwidth]{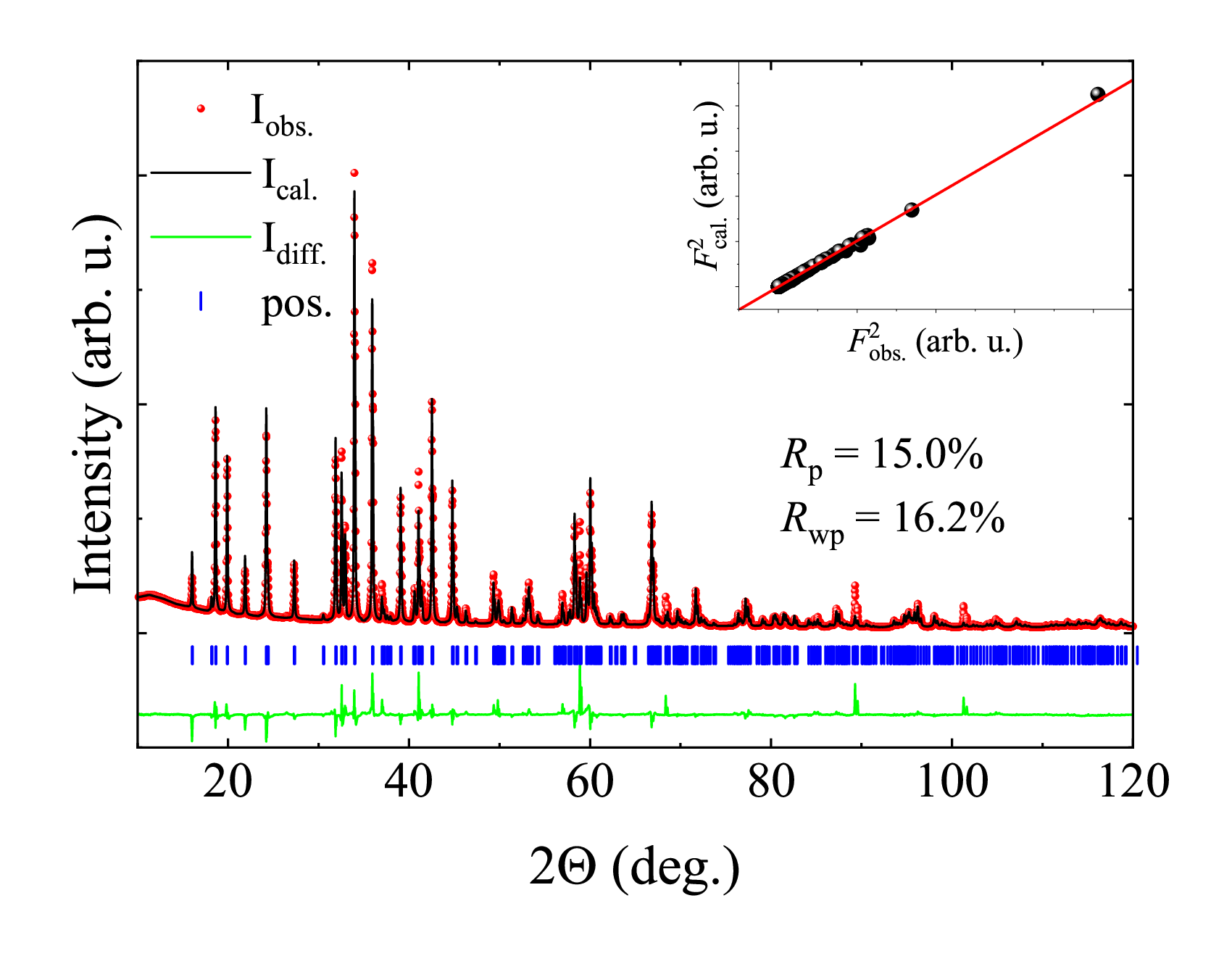}
\caption{\label{xrd} Rietveld refinement of the powder X-ray diffraction pattern for \pmao. The inset shows the $F_{obs.}^2$ vs. $F_{cal.}^2$ from single crystal refinement. The straight line is a guide to the eye.}
\end{figure}

The powder X-ray diffraction pattern for the crushed single crystals is shown in Fig. \ref{xrd}. Due to the strong preferred orientation related to the two-dimensional character of the sample, the final agreement factors are relatively high even when the preferred orientation was considered during the refinement using the Fullprof package. However, the key information is that no additional peaks were observed from this measurement, indicating an impurity-free phase within our experimental resolution.

\subsection{X-ray Photoelectron Spectroscopy}

\begin{figure}[htbp]
\centering
\includegraphics[width=0.5\textwidth]{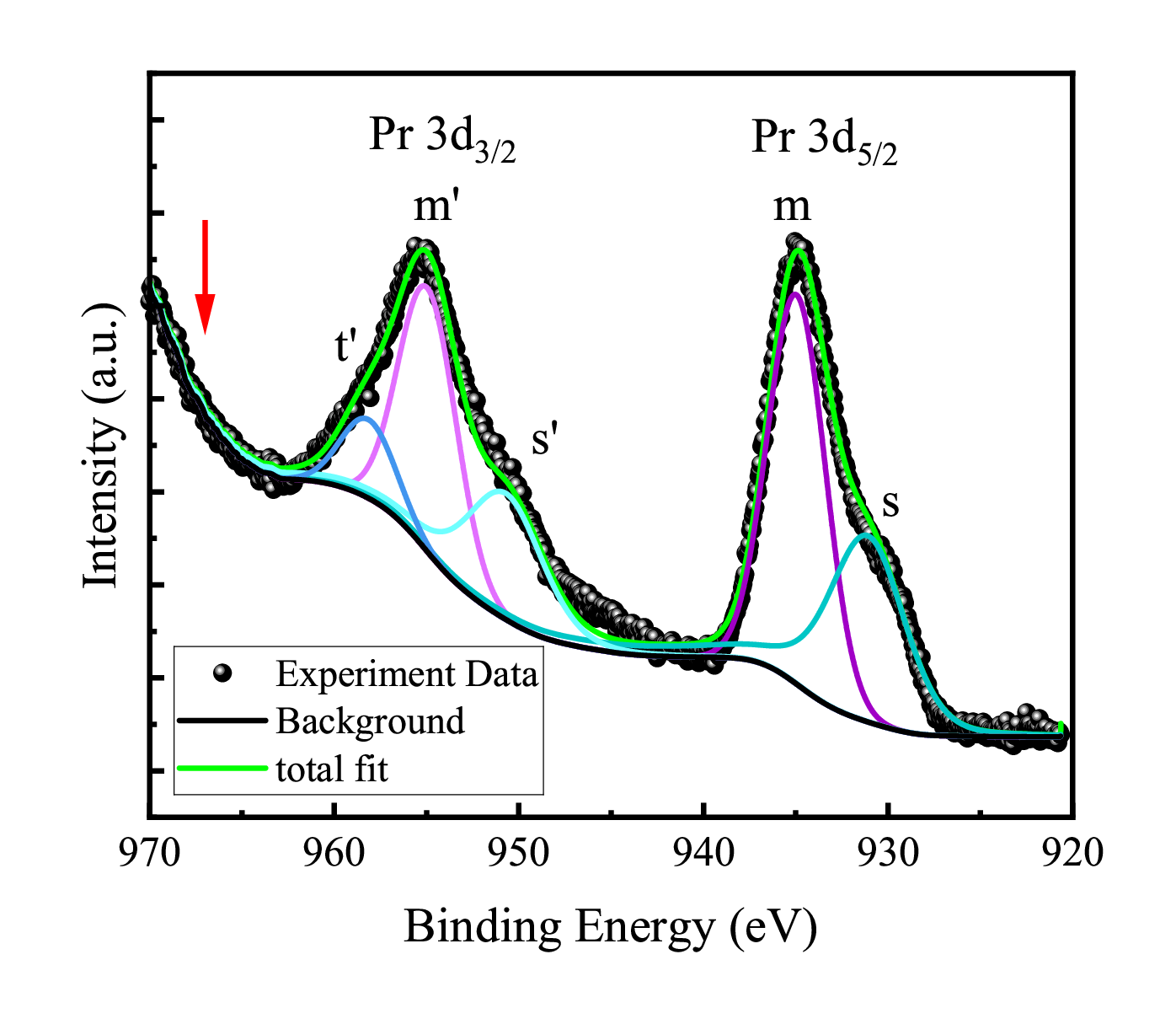}% Here is how to import EPS art
\caption{\label{xps}$\rm{Pr}$ 3$d$ X-ray photoelectron spectroscopy for \pmao\ single crystal. The red arrow indicates the expected peak position for the Pr$^{4+}$ state.}
\end{figure}

Due to the fact that Pr ion can display different valence states in oxides, we performed X-ray photoelectron spectroscopy (XPS) measurements to investigate the valence of Pr in our single crystals. The measurements were carried out on an XPS spectrometer (Thermo Fisher ESCALAB XI+) equipped with a monochromated Al  $K_{\alpha}$ X-ray source. The spectra are fitted using the Avantage software.

Previous studies have demonstrated that Pr 3$d$ XPS can clearly distinguish between $\rm{Pr}^{4+}$ and $\rm{Pr}^{3+}$ \cite{Bianconi1988,Ogasawara1991,holland1992,gurgul2013}. As shown in Fig. \ref{xps}, the spectrum exhibits two main peaks (m and m$'$) which are due to the spin-orbit split 3d$_{3/2}$ and 3d$_{5/2}$ core holes. In addition, several satellites (s, s$'$ and t$'$) can be observed due to the multiplet effect. All these features are reminiscent of the results for $\rm{Pr}_2\rm{O}_3$ \cite{Pr1995}. Moreover, no marker peak for $\rm{Pr}^{4+}$ ($\sim 967.0$ eV) was observed. Thus, we can safely conclude that a Pr$^{3+}$ state is formed in our single crystal.

\subsection{Electron spin resonance}

\begin{figure}
  \centering
  \includegraphics[width=1\columnwidth]{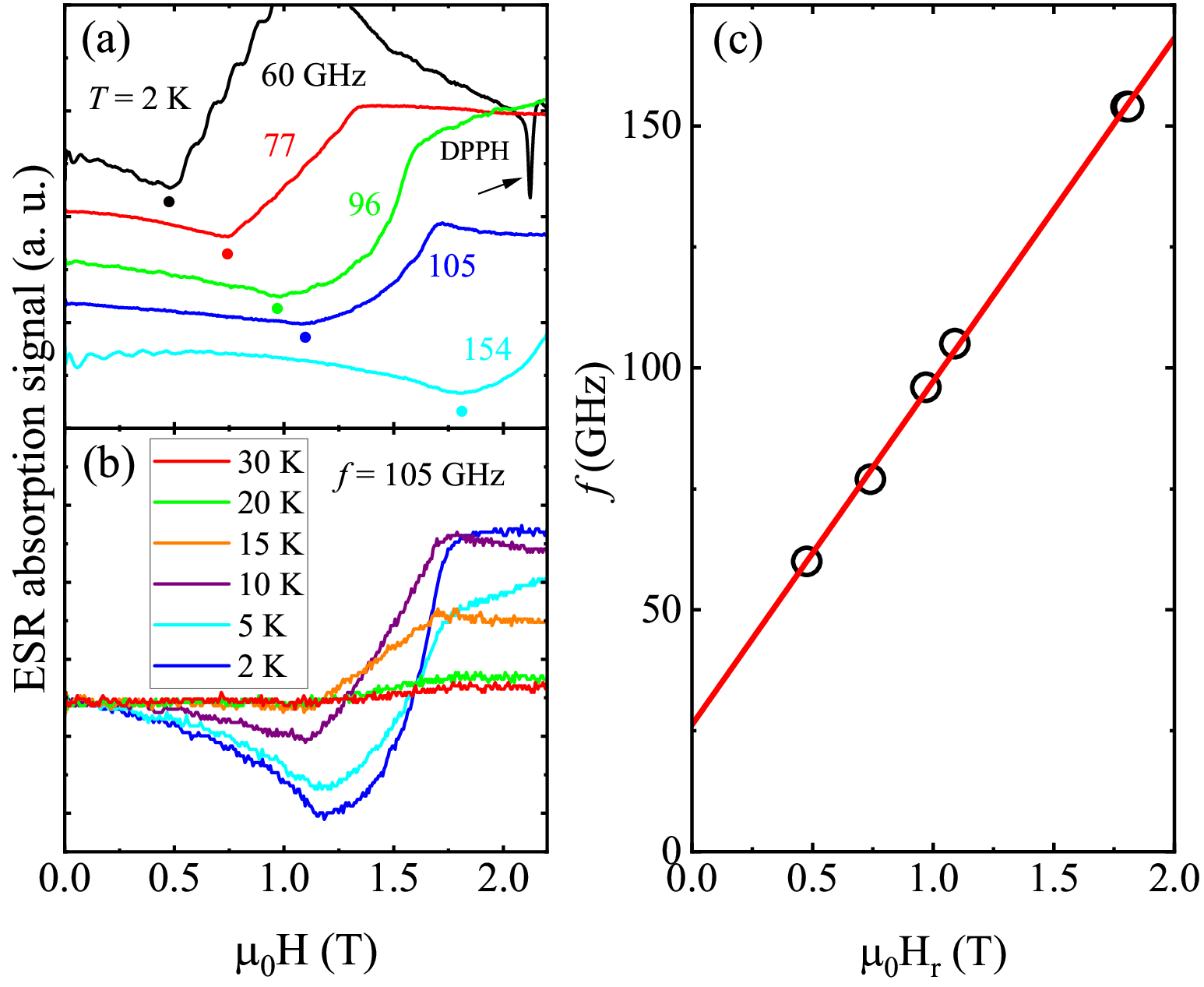}\\
  \caption{(a) Electron spin resonance spectroscopy for \pmao\ measured at 2 K with magnetic field applied along the \textit{c} axis. The small dots under each line indicate the resonant field $H_r$ positions. (b) The \textit{f} = 105 GHz spectra measured at various temperatures. (c) Magnetic field dependence of the ESR absorption frequency. The red line is a linear fit, see the text for details.}\label{esr}
\end{figure}

The high-field electron spin resonance (ESR) spectra were collected in the field-increasing process in a frequency range of 60 to 154 GHz at Wuhan National High Magnetic Field Centre (WHMFC). The resonance line of standard sample DPPH with g = 2.0 was used for a field marker. The absorption spectra measured at different frequencies are shown in Fig. \ref{esr}(a), from which the resonant field, $H_r$, can be identified. As can be seen from Fig. \ref{esr}(b), the absorption becomes more and
more indiscernible at high temperatures, suggesting that we
are probing the ground state. Moreover, as shown in Fig. \ref{esr}(c), the absorption frequency depends linearly on $H_r$ such that $f = f_0 + g_c\mu_\mathrm{B}\mu_0H_r/h$, where $\mu_B$ is the Bohr magneton, $h$ is the Planck constant. The best fit yields the $g_c$ value of 5.1 and $f_0$ of 26.2 GHz. The nonzero $f_0$ indicates the presence of a zero-field gap of $\sim$0.1 meV.

\subsection{Crystal-electric-field analyses}

\begin{table*}
\caption{CEF eigenvectors and eigenvalues for Pr$^{3+}$ at the 2\textit{d} site.}
%\begin{ruledtabular}
\begin{tabular}{c|ccccccccc}
E (meV) &$|-4\rangle$ & $|-3\rangle$ & $|-2\rangle$ & $|-1\rangle$ & $|0\rangle$ & $|1\rangle$ & $|2\rangle$ & $|3\rangle$ & $|4\rangle$ \tabularnewline
 \hline
0.000 & -0.9588 & 0.0 & 0.0 & 0.0 & 0.0 & 0.0 & -0.2842 & 0.0 & 0.0 \tabularnewline
0.000 & 0.0 & 0.0 & -0.2842 & 0.0 & 0.0 & 0.0 & 0.0 & 0.0 & -0.9588 \tabularnewline
19.949 & 0.0 & 0.7071 & 0.0 & 0.0 & 0.0 & 0.0 & 0.0 & 0.7071 & 0.0 \tabularnewline
42.086 & -0.2842 & 0.0 & 0.0 & 0.0 & 0.0 & 0.0 & 0.9588 & 0.0 & 0.0 \tabularnewline
42.086 & 0.0 & 0.0 & -0.9588 & 0.0 & 0.0 & 0.0 & 0.0 & 0.0 & 0.2842 \tabularnewline
43.287 & 0.0 & 0.0 & 0.0 & -1.0 & 0.0 & 0.0 & 0.0 & 0.0 & 0.0 \tabularnewline
43.287 & 0.0 & 0.0 & 0.0 & 0.0 & 0.0 & -1.0 & 0.0 & 0.0 & 0.0 \tabularnewline
46.241 & 0.0 & 0.0 & 0.0 & 0.0 & -1.0 & 0.0 & 0.0 & 0.0 & 0.0 \tabularnewline
50.291 & 0.0 & -0.7071 & 0.0 & 0.0 & 0.0 & 0.0 & 0.0 & 0.7071 & 0.0 \tabularnewline
\end{tabular}
%\end{ruledtabular}
\label{EV}
\end{table*}

As described in the main text, the CEF eigenvalues and eigenvectors extracted from the fitting to the $\chi_c^{-1}(T)$ data are shown in Tab. \ref{EV}. The calculated $g_c$ from the ground state doublet amounts to 5.6, which is consistent with the ESR measurement. Since Pr$^{3+}$ is a non-Kramers ion, the ground state doublet is not protected by the time reversal symmetry, and can be split by further lattice distortions, which is plausible regarding the site disorder at the Pr site. The zero field splitting observed by ESR supports this speculation. However, the splitting is very small ($\sim$ 0.1 meV), the ground state can still be viewed as a quasidoublet, as observed in many related non-Kramers systems \cite{Wen2017,Li2018}. In this case, the longitudinal spin component behaves as the magnetic dipole moment, while the transverse component like multipoles \cite{Shen2019,chen2019}.

%\bibliography{cobaltate}

%For data citations of datasets uploaded to e.g. \emph{figshare}, please use the \verb|howpublished| option in the bib entry to specify the platform and the link, as in the \verb|Hao:gidmaps:2014| example in the sample bibliography file.

\end{document}